\documentclass[epj]{svjour}
\usepackage{amsmath,amssymb,amsfonts}
\usepackage{amscd}
\usepackage{graphics}
\begin{document}
\title{Nonequilibrium Phase Transitions into Absorbing States}
\subtitle{Focused around the pair contact process with diffusion}
\author{Su-Chan Park\inst{1} \and Hyunggyu Park\inst{2}
}                     
%
%
\institute{Institut f\"ur Theoretische Physik, Universit\"at zu K\"oln, Z\"ulpicher Str. 77, 50937 K\"oln, Germany, \email{psc@thp.uni-koeln.de} \and School of Physics, Korea Institute for Advanced Study, Seoul 130-722,
Korea, \email{hgpark@kias.re.kr}}
\date{Received: date / Revised version: date}
%
\abstract{Systems with absorbing (trapped) states may exhibit a
nonequilibrium phase transition from a noise-free inactive phase
into an ever-lasting active phase. We briefly review the absorbing
critical phenomena and universality classes, and discuss over the
controversial issues on the pair contact process with diffusion
(PCPD). Two different approaches are proposed to clarify its
universality issue, which unveil strong evidences that the PCPD
belongs to a new universality class other than the directed
percolation class.
} 
\maketitle
\section{\label{Sec:intro}Introduction}
The absorbing phase transition (APT) which may occur in a system
with absorbing states has been an active research topic in
nonequilibrium statistical mechanics with possible applications to
a wide range of areas in physics, chemistry, biology, and
sociology~\cite{MD1999,H2000,O2004,L2004}. Microscopic states into which
probability is collected are called absorbing.  Once a system is
trapped into absorbing states, it cannot escape. Absorbing states
sometimes form an absorbing space of many states around which a
system wanders forever, but does not escape out of the absorbing
space. Such an example is the state with only one particle
diffusing in physical space for the pair contact process with
diffusion (PCPD); see Sect.~\ref{Sec:PCPD}.

In the study of the APT, two different quantities may serve as
order parameters\footnote{Although the APT has nothing to do with
the order-disorder transition, the jargon ``order parameter'' has
been used to name the indicator of the phase transition which
takes zero in one phase and non zero in the other.}. One is the
steady state density of the activity (outside of the absorbing
states) in the thermodynamic limit\footnote{For systems with
finite volume of the configurational state space, the presence of
the absorbing state always lets a system fall into it
eventually~\cite{vK1997}. So, as in the equilibrium statistical
mechanics, the thermodynamic limit (infinite extension of the
configurational state space) is indispensable to study a
nontrivial APT.}, say $\rho$ whose definition varies from system
to system, and the other is the survival probability, say $P_s$,
with which a system  does not get trapped into absorbing states
forever. Since a finite number of activity (or local activity)
means zero activity density in the thermodynamic limit but non
zero survival probability, $\rho=0$ in general does not imply $P_s
= 0$ although the reverse is always true. In a sense, $\rho$
($P_s$) is the order parameter for the ``macroscopic''
(``microscopic''\footnote{The term ``microscopic'' is employed
because the survival probability has been mainly analyzed in
practice when the activity is initially localized in an infinite
lattice. In this case, the macroscopic density $\rho$ remains zero
for all time by definition though the number of activity may
increase indefinitely with finite probability. Hence, the
meaningful activity density $\rho$ in the thermodynamic limit
should be examined with a finite initial density.}) APT. Hence,
these two quantities might pinpoint different transition points in
principle, see Sect.~\ref{Sec:PCPD} for an example. However, with
proper initial conditions and definitions for each order
parameter, these transitions points may coincide in general.

Besides the difference mentioned above, nature of the phase
transition described by these order parameters can be different
even if they locate the same critical point. To clarify this
point, consider the branching process with spontaneous death
($A\rightarrow 2A$ with rate $\sigma$ and $A\rightarrow 0$ with
rate $\lambda$). The absorbing state of this model is a state
without $A$. This problem is the linear one step process in
Ref.~\cite{vK1997}, which can be solved easily and one may find
the survival probability as
\begin{equation}
P_s = \left \{ \begin{array}{cc} 0 & \text{ if } \lambda \ge \sigma\\
\displaystyle \frac{\sigma-\lambda}{\sigma} & \text{ if } \lambda \le \sigma
\end{array} \right ..
\end{equation}
On the other hand, the mean number of particles (activity), say
$\langle n \rangle$, behaves as
\begin{equation}
\frac{\partial \langle n \rangle}{d t} = (\sigma - \lambda) \langle n \rangle
\Rightarrow
\langle n \rangle = n_0 e^{(\sigma-\lambda)t},
\end{equation}
with the initial value $n_0$. Both quantities are singular at
$\lambda = \sigma$, but $P_s$ increases continuously unlike
$\langle n \rangle$ which shows a sharp jump at the transition
point. Hence in the study of the APT, $P_s$ and $\rho$ should be
studied independently and both quantities are important in
understanding the APT.

In what follows, however, we restrict ourselves to the study of
the ``macroscopic'' APT, or the scaling behavior of $\rho$ and
quantities related to it like the correlation functions. Like
equilibrium phase transitions, the APT is characterized and
classified by the critical exponents; the order parameter exponent
$\beta$, correlation length exponent $\nu_\perp$, relaxation time
exponent $\nu_\|$, and so on, which are defined as
\begin{equation}
\begin{aligned}
\rho \sim ( p_c -p )^\beta,\; \xi \sim |p-p_c|^{-\nu_\perp}, \;
\tau \sim |p-p_c|^{-\nu_\|},
\end{aligned}
\end{equation}
where $p$ is the external tuning parameter with $p_c$ to be the
critical point,  $\xi$ is the correlation length, and $\tau$ is
the relaxation time. One can define other critical exponents but
with the aid of the scaling ansatz most of critical exponents can
be deduced from the above exponents~\cite{GT1979}. For example,
the density decays as $t^{-\delta}$ at criticality with $\delta =
\beta/\nu_\|$~\cite{GT1979}.

To categorize universality classes according to the critical
exponents and to understand what properties combine different
systems into the same universality class are the main goals in
this field. Some understanding has emerged from the numerical and
analytical studies.  Section~\ref{Sec:univ} briefly summarizes the
well-established universality classes, such as the directed
percolation (DP), the directed Ising (DI), the parity conserving
(PC) classes and so forth. The last 10 years have witnessed intensive
discussion and hot debate on a simple but very elusive interacting
particle system, the PCPD. In Sect.~\ref{Sec:PCPD}, we will
critically review on the issue regarding its universality class.
To settle the controversy, two different approaches have been
proposed by the authors, which is the subject of
Sect.~\ref{Sec:resolve}. We draw a conclusion in
Sect.~\ref{Sec:con}.

\section{\label{Sec:univ} Universality classes}
The simplest non-trivial model which shows an APT might be the
contact process (CP) which is an interacting hard core particle
system on a $d$ dimensional lattice with the creation of a
particle by a neighboring particle and the spontaneous
annihilation ($A\rightarrow 2A$, $A\rightarrow 0$)~\cite{H1974}.
The particle vacuum state is the only absorbing state in which the
system cannot escape by the prescribed rules. Hence the order
parameter is the density of occupied sites (or particles). The CP
shares the critical behavior with the directed percolation (DP)
the preferred direction of which is interpreted as the time
direction of the CP and the open channel of which as a
particle~\cite{H2000}.  The DP has the rapidity-reversal (or time-reversal)
symmetry~\cite{GT1979,J1981} which associates the microscopic APT with the macroscopic one,
that is, which renders $P_s$ to scale equivalently to $\rho$.
A nice illustration of the connection in the context of the bond directed percolation can
be found in Ref.~\cite{H2000}; see also Ref.~\cite{L2004}.

After extensive numerical studies regarding
the universality class of the APT, it has been
conjectured\footnote{This statement is termed as the ``DP
conjecture''.} that the APT model with a \textit{single} absorbing
state should belong to the DP class if symmetry or conservation is
not involved~\cite{J1981,G1982,GLB1989}.
The robustness of the DP class extends to the systems with
infinitely many absorbing states like the pair contact process
(PCP)~\cite{J1993}, at least in its stationary property.  In the
PCP, the creation and annihilation of particles are only mediated
by two particles which form a nearest neighbor pair on a lattice
($2A\rightarrow 3A$, $2A\rightarrow 0$). So any configuration
devoid of a pair is an absorbing state and the number of absorbing
states increases exponentially with system size. Clearly, the
order parameter should be the pair density not the particle
density (auxiliary field density) which is nonzero at stationarity
irrespective of the phase. The robustness of the DP class suggests
that the absence of symmetry or conservation of the order
parameter should render the system to belong to the DP class,
irrespective of the existence of the auxiliary field associated
with infinitely many absorbing states. Although the stationary
property of the PCP conforms with the DP scaling, there is still
vivid discussion regarding the dynamic scaling of its
spreading~\cite{MGD1998,JDH2003,D1996,GCR1997}, which is beyond
the scope of this paper.

Other universality classes have been found by adding symmetry or
conservation in dynamics. The directed Ising (DI) class involves
the $Z_2$ (Ising) symmetry in dynamics, the evolution operator of
which is invariant under the $Z_2$ symmetry
operation~\cite{KP1994,PKP1995,HKPP1998}. Naturally, the typical
DI systems include two equivalent absorbing states. The DI scaling
also applies to systems with two equivalent groups of multiple
absorbing states~\cite{HP1999,MM1999,PP2001}. The conservation of
the particle number of modulo 2 reveals another universality class
called as the parity conserving (PC) class~\cite{G1989,TT1992},
which coincides with the DI class in one dimension. In higher
dimensions, both classes can be described by the trivial
mean-field theory. Recently another universality class
(generalized voter class) has been examined, which also coincide
with the DI class in one dimension~\cite{HCDM2005}. It is worthy
to note that the DI (and PC) class returns to the DP class
immediately with the introduction of a symmetry breaking field or
a conservation breaking
dynamics~\cite{PP1995,KP1995,BB1996,H1997}.

There had been an attempt to find a new universality class by
studying models with higher symmetry than $Z_2$ or mod($q$)
conservation with $q>2$. However, all models so far show a trivial
critical behavior even in one dimension, in that the absorbing
phase is found to be always unstable against the dynamics
increasing activity of the order parameter.

All the systems explained up to now can be described by the single
component order parameter. Richer behavior is anticipated when
multi species are involved. For example, the interaction of the
order parameter with a conserved field triggers a different type
of universal behavior depending on the activity of the conserved
field~\cite{KSS1989,WOH1998}. In fact, any system involving
multi-particle reactions can be interpreted as a multi-species
particle system. One can map the PCP to the multi-species model by
identifying a pair as a particle of one kind and a single isolated
particle as a particle of another kind. More general cases will be
discussed in subsequent sections.

\section{\label{Sec:PCPD}Pair contact process with diffusion}
The pair contact process with diffusion (PCPD) is an extended
model of the PCP with hopping of particles allowed for. To be
specific, the dynamics of the PCPD in $d$ dimensions consists of
hopping, pair annihilation, and  creation by a pair, which is
symbolically summarized as
\begin{subequations}
\label{Eq:PCPDrule}
\begin{eqnarray}
\label{Eq:hop}
\left .
\begin{array}{c}
A \emptyset \rightarrow \emptyset A\\
 \emptyset A\rightarrow A\emptyset
\end{array} \right \} \text{ with rate } \frac{D}{d},\\
\label{Eq:ann}
A A \rightarrow \emptyset \emptyset \text{ with rate } \frac{p}{d},\\
\label{Eq:cre}
\left .
\begin{array}{c}
AA \emptyset\\
 \emptyset AA
\end{array} \right \} \rightarrow AAA\text{ with rate } \frac{1-p}{2d},
\end{eqnarray}
\end{subequations}
where $A$ ($\emptyset$) stands for an occupied (a vacant) site on
a lattice and $0 \le p \le 1$.  The PCP corresponds to the above
rules with $D=0$ which make any configuration without two
particles in a row (a pair) absorbing. Due to the diffusion,
however, a state without a pair but many isolated particles is not
absorbing any longer. Only both the particle vacuum and the state
with only one particle in the whole system are absorbing. Since
the particle density of both absorbing states of the PCPD is zero
in the thermodynamic limit, it can play the role of the order
parameter in contrast to the PCP case. Needless to say, the pair
density may also serve as an order parameter.

Due to the lack of a process to eliminate a single particle (no
single particle reaction) without particle collisions, the {\em
conventional} survival probability and the density might locate
different transition points. To elucidate, consider the PCPD
in higher dimensions than 2. If initially two particles are
located somewhere in an infinite lattice (just outside of the
absorbing space), the survival probability is always finite
because of the nonrecurrence of the random walk even for the case
of $p= 1$. That is, the survival probability predicts the absence
of the ``microscopic'' absorbing phase. However, the macroscopic
critical point should be located at finite $p$. The reason is as
follows: The mean density $\rho$ of particles in $d$ dimensions
satisfies the (exact) equation,
\begin{equation}
\frac{ d \rho}{d t} = (1 - 3 p) \rho_p - (1-p) \rho_t,
\label{Eq:pcpd_rho}
\end{equation}
where $\rho_p$ ($\rho_t$) means the pair (triplet) density. If
$p>\frac{1}{3}$, the steady state value of the pair and triplet
density should be zero, which is clear by Eq.~\eqref{Eq:pcpd_rho}.
If $\rho$ were not zero in the steady state, macroscopic number of
pairs should be formed by the diffusion, which is contradictory to
the observation that $\rho_p=0$. Hence $\rho$ should approach to
zero if $p>\frac{1}{3}$ and the critical point should be not
larger than $\frac{1}{3}$ and the ``macroscopic'' absorbing phase
is present in any dimension\footnote{To resolve the difference
between the microscopic and macroscopic APT,  a {\em new}
definition of survival of the system has been suggested: The
system without a pair is considered as (temporarily) absorbing.
With this definition, it has been shown that two order parameters
exhibit the APT's at the same transition point at least in one
dimension~\cite{NP2004,DM2002}.}. In the above argument, we assume the
existence of the steady state even in the thermodynamic limit.

As the above consideration reveals, the single-particle diffusion
plays a crucial role in changing the nature of the conventional
microscopic APT, which has nothing to do with the DP for higher
dimensions than 2. However, this does not resolve the controversy
regarding the universality class of the PCPD. First, the
difference of the microscopic APT does not guarantee that of the
macroscopic one. A good example is the PCP whose macroscopic APT
is characterized by the DP scaling but whose microscopic APT is
known to be nonuniversal~\cite{MGD1998,JDH2003,D1996,GCR1997}.
Second, the main issue is not any dimensional PCPD but one
dimensional PCPD where it is not fully clear that the microscopic
APT is equivalent to the macroscopic APT~\cite{NP2004,DM2002}.

In fact, the difference between the PCPD and the DP is well appreciated
for two or more dimensional systems.  First consider the mean field equation.
The mean field equation for the PCPD can be found by replacing
$\rho_p$ and $\rho_t$ with $\rho^2$ and $\rho^3$, respectively
in Eq.~\eqref{Eq:pcpd_rho} which reads
\begin{equation}
\frac{\partial \rho}{\partial t} = (1- 3 p) \rho^2 - (1-p) \rho^3.
\end{equation}
The mean field critical exponents are $\beta = 1$ and $\delta =
\beta/\nu_\| =\frac{1}{2}$ \cite{HH2004,PHP2005b}, which are
different from those of DP ($\beta=\nu_\|=1$)~\cite{H2000}. Since
the upper critical dimension of the PCPD is believed to be
$2$~\cite{OMS2002}, for most physically relevant cases ($d\ge 2$)
the PCPD does not belong to the DP class.

For the one dimensional PCPD, however, the numerically estimated
critical exponents of the PCPD are so similar to those of the DP
that the possibility for the PCPD to belong to the DP class has
been raised~\cite{BC2003,H2006}. Interestingly, the critical
exponent\footnote{From now on, by critical indices without
subscript are always meant those of the PCPD.}  $\delta$ which
describes the density decay with time at criticality has floated
from $\simeq 0.28$~\cite{CHS2001} to less than
$0.185$~\cite{H2006} with time, which is due to the strong
corrections to scaling. For comparison, the numerical value of
$\delta_\textrm{DP}$ is $\simeq 0.15946$~\cite{J1996}.

An argument in favor of the DP scenario was suggested by
Hinrichsen~\cite{H2006}. The starting point of the argument is the
numerical observation that the dynamic exponent $z$ is smaller
than 2 which is the dynamic exponent of the random walk. Hence, if
coarse graining is performed according to the PCPD dynamic
exponent, the diffusing isolated particles will stop moving
asymptotically and the long time behavior of the PCPD should be
same as that of the PCP which belongs to the DP.

We would like to make some comments as to this argument. To begin,
it is not at all clear why the dynamics of the wandering isolated
particles is decoupled asymptotically from that of the active
clusters. If not, the dynamic scaling of the isolated particles
should be affected by the complex environmental geometry of active
clusters. Therefore there is no ground for the belief that the
diffusion of isolated particles remains governed by the random
walk dynamic exponent $z_\text{RW}=2$. If the system scales in one
way as a whole (not decoupled), the coarse graining argument does
not lead to the zero diffusion constant of the isolated particles.
Second, the Hinrichsen's argument set the upper bound for the
dynamic exponent $z\le 2$ in order to be self-consistent. Based on
this, one can comment on the two dimensional PCPD which seems
definitely not in the DP class. This leaves us only one option to
take $z=2$ for the two (and higher) dimensional PCPD even without
any logarithmic correction. However, the upper critical dimension
of the PCPD is believed to be 2, where the logarithmic correction
is expected and the numerical results seem to support its
presence~\cite{OMS2002}. Therefore, even if one may accept the
decoupling of two different fields in the PCPD, his argument seems
not working at two dimensions.

\begin{figure}[t]
\resizebox{\columnwidth}{!}{
\includegraphics{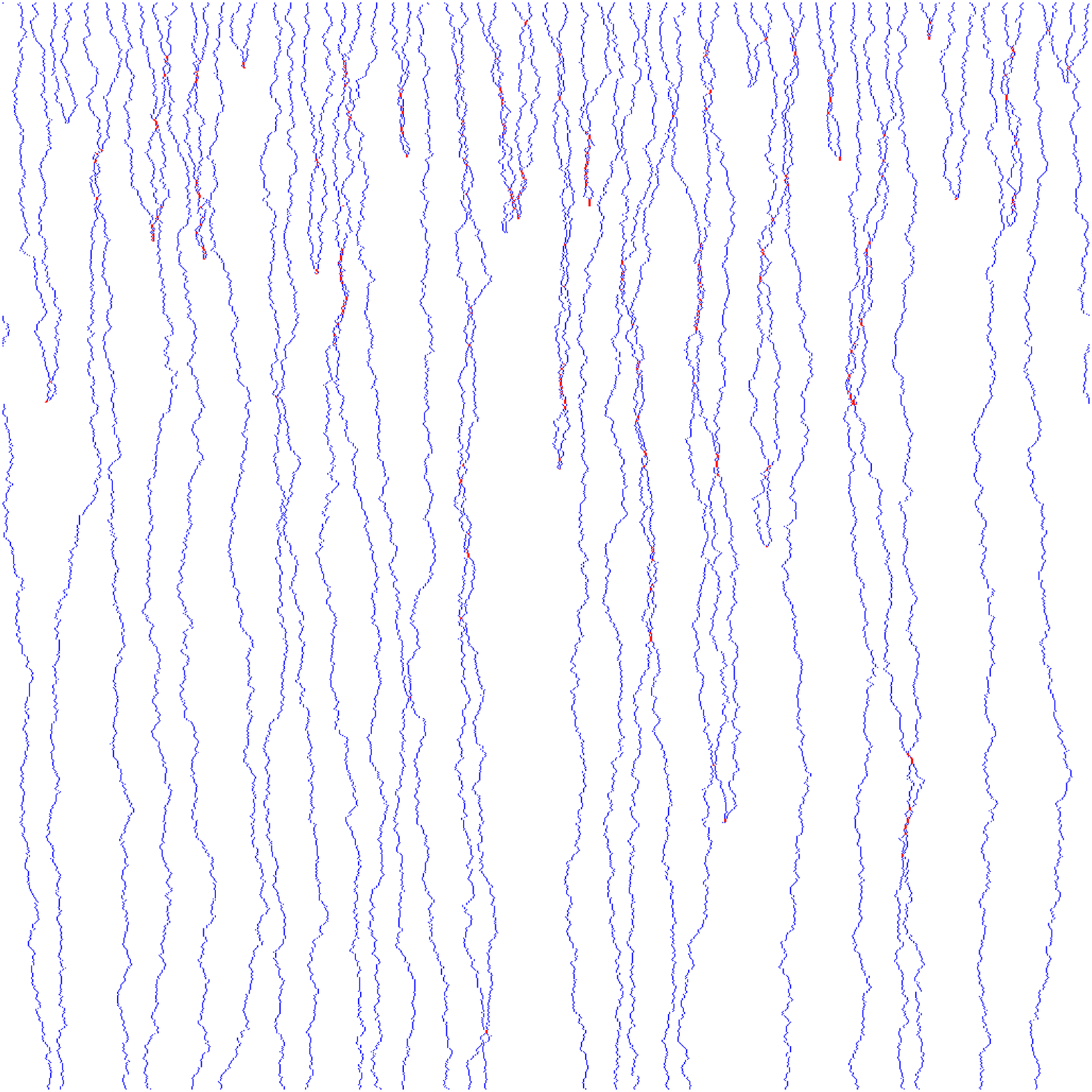}
} \caption{\label{Fig:tp12} (Color online) Space-time configuration of the
\texttt{tp12} at criticality with a sparse initial condition. The
time flows from top to bottom and the cross section by a
horizontal line is the space configuration at specified time. The
periodic boundary conditions are employed. Effective lack of the
memory (see the text) means the seemingly absence of the branching
process $3A\rightarrow 4A$ due to the low probability of forming a
triple before a pair is annihilated.}
\end{figure}

Regardless of the universality issue, one can ask why the PCPD has
such strong corrections to scaling which are the main obstacle in
numerical study. The long term memory effect was suggested as a
possible origin of the strong corrections to scaling though it is
not clearly answered how the finite mean life time from the life
time distribution $P(\tau) \sim \tau^{-2.2}$ can trigger such
strong corrections to scaling or even a new scaling deviated from
the DP~\cite{NP2004}. Only a possible scenario was contemplated
that the interacting theory may force the system to flow into a
non-DP fixed point even with finite mean life time, reflecting on
the similar result found in the L{\' e}vy flight DP
systems~\cite{JOvWH1999,HH1999}. Due to the fact that the
memoryless case of the generalized PCPD clearly shows the DP
scaling~\cite{NP2004}, it is certain that the memory effect plays
an important role in determining the universality class of the
PCPD. Another reason of believing the role of the memory comes
from the model \texttt{tp12} (for the definition, see below)
and similar varieties of models with
rules $nA \rightarrow (n+m) A$ and $l A \rightarrow (l - k) A$
with $n>l$ and $m,k>0$ which are numerically shown to belong to
the DP~\cite{KC2003}.  Same as the PCPD, the \texttt{tp12}
($n=3,l=2,m=1,k=2$) with zero diffusion constant has infinitely
many absorbing states and belongs to the DP class~\cite{PP2007}.
In contrast to the PCPD ($n=2,l=2,m=1,k=2$), the diffusion turns
out not to affect the universality class of the \texttt{tp12}. In
Ref.~\cite{PP2007}, it is argued that the effective lack of the
memory in the \texttt{tp12}, which is clearly seen from the
space-time configuration Fig.~\ref{Fig:tp12}, renders any model
with $n>l$ to belong to the DP class.

 Another complicated feature of the PCPD has
arisen in the analytical study. The formulation of the field
theory based on the single field turned out impossible and it is
argued that the nonperturbative treatment still cannot cure this
failure~\cite{JvWDT2004}. The lesson from the study is that at
least two independent field should be included in the proper
action to understand the PCPD in the field theoretical framework.
The proper field theory which is local in space and
time\footnote{The effective field theory might be described by a
single field with nonlocal interaction, which in principle can be
achieved by integrating out some of the fields in a local
action.}, if exists, should take two independent fields into
account. In principle, one can write down the Langevin equation
equivalent to the PCPD using two independent fields~\cite{P2006},
but the proper analytical tool does not seem to be at hand.

\section{\label{Sec:resolve}How to tackle the problem}
The one dimensional PCPD seems very difficult, if not impossible,
to tackle directly. In the numerical front, the strong corrections
to scaling prohibits researchers from measuring the critical
exponents accurately\footnote{The most recent numerical studies
still do not provide a conclusive evidence~\cite{OD06,KK07}.}. In
the analytical front, the failure of the field theory by the
single field requires an ingenious treatment of the problem.

In this section, we will try to convince readers that the PCPD does not
belong to the DP class. First, introducing the bias, we will argue
that the failure of the field  theory by the single field
is also generic even in one dimension. Second, studying the crossover model
from the PCPD to the DP by introducing the single particle dynamics,
the difference of these two classes will be clarified.
\subsection{\label{Sec:DPCPD}Effect of Biased Diffusion}
As discussed at the end of Sect.~\ref{Sec:PCPD}, the PCPD is supposed to
be described by two independent ``elementary excitations'', that is,
the isolated particle-field and the pair-field.

Even if we find the proper field theory for the PCPD, it might not
resolve the controversy on the one dimensional PCPD. For example,
what if the fixed point of the PCPD turns out not to be reachable
by the perturbation expansion to the one dimensional system just
as that of the branching annihilating walk (BAW) with even number
of offspring~\cite{CT1996}? However, before worrying about the
above scenario, we must take a first step to answer the most
elementary question whether it is absolutely necessary to employ
two independent fields in the one dimensional PCPD, in contrast to
the Reggeon field theory of the DP which needs a single field. The
answer may not resolve the controversy definitely, but should be
regarded as a big step in understanding the difference between the
PCPD and the DP class.

Recently, the present authors have suggested to check if the one
dimensional PCPD should have two independent relevant fields via
adding a bias in diffusion~\cite{PHP2005a,PHP2005b}. We got some
hints from the study of the two species annihilation model $A+B
\rightarrow \emptyset$ which shows different decaying behavior in
the presence of the relative  bias between $A$ and $B$ from that
without bias~\cite{KR1984,KR1985}.

\begin{figure}[t]
\resizebox{\columnwidth}{!}{
\includegraphics{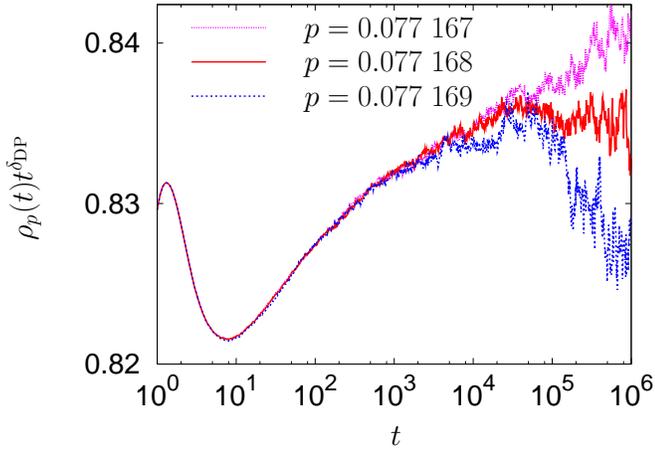}
}
\caption{\label{Fig:DPCP} (Color online) Semi-logarithmic plot of the pair density multiplied
by $t^{\delta_\text{DP}}$ with $\delta_\text{DP} = 0.1594$ vs time
for the PCP with biased branching.
In the active (absorbing) phase, the curve veers up (down) and at criticality
the flat straight line is expected, which renders the estimation of the
critical point as $p_c = 0.077~168(1)$ with error in the last digit.
}
\end{figure}
Since the DP has only one relevant field, the bias should be gauged
away by the Galilean transformation.  Besides, lots of complicated models
which belong to the DP class also turn out to be robust against the bias
\cite{PHP2005a,PP2007}.
Implementing the relative bias between relevant fields and observing
how the system responds can be a litmus test to check numerically
whether the single field is enough to describe the system.
In this context, we introduced the driven pair contact process
with diffusion (DPCPD) and studied it \cite{PHP2005a,PHP2005b}.

Before delving into the DPCPD, we should emphasize that the
presence of two ``independent'' fields is not sufficient to
observe the change of critical behavior by the bias. Although the
PCP is described by two independent fields, we expect that the
branching bias which triggers the relative bias between pairs and
isolated particles does not alter the critical behavior because
the PCP belongs to the DP class. In Fig. \ref{Fig:DPCP}, we showed
how the pair density behaves around the critical point of the PCP
with biased branching. As in the PCP, $p$ is the pair annihilation
probability and $1-p$ is the creation probability. The branching
bias is given such that a created particle is always placed at the
right neighbor of the pair once that site is vacant. Figure
\ref{Fig:DPCP} locates the critical point $p_c$ as $p_c =
0.077~168(1)$ by exploiting the power law behavior of the pair
density, say $\rho_p$, at criticality. As can be seen, the
critical decay exponent  still takes the DP value
($\delta_\text{DP}$), which implies that the relative bias is
irrelevant in the PCP.

Now let us continue the discussion on the DPCPD. In
Ref.~\cite{PHP2005a}, the DPCPD shows clear distinction from both
the DP and the PCPD scaling in one dimension, which was confirmed
again by the cluster approximation along with the coherent anomaly
method analysis~\cite{PHP2005b}. This result is not expected if
there is only one relevant field in the PCPD. Hence the field
theory of the PCPD should be different from the Reggeon field
theory even in one dimension, which provides a strong evidence of
a non-DP scaling in the PCPD.

Another interesting feature of the DPCPD is the dimensional
reduction. The one dimensional DPCPD is numerically found to have
the same critical behavior as the two dimensional
PCPD~\cite{PHP2005a}. The dimensional reduction by the biased
diffusion is actually reported in different areas in
nonequilibrium statistical physics. One example is the two species
annihilation model, $A + B\rightarrow 0$. When there is (no)
relative bias between $A$ and $B$, the density decays as
$t^{-(d+1)/4}$ ($t^{-d/4}$) when $d<3$ ($d<4$) and as $t^{-1}$
when $d>3$ ($d>4$)~\cite{KR1984,KR1985}. Another example can be
found in the study of the self organized criticality. It is
rigorously proved that the upper critical dimension of the
directed sand pile model is 3 unlike its undirected version whose
upper critical dimension is 4~\cite{DR1989}.

\subsection{\label{Sec:crossover} Learning from crossover scaling}
Another evidence that the PCPD does not belong to the DP was provided
by the study of the crossover from the PCPD to the DP~\cite{PP2006}.
Along with the dynamics in Eqs.~\eqref{Eq:PCPDrule}, we introduce
the single particle annihilation/creation
\begin{equation}
\begin{CD}
A \stackrel{wq}{\longrightarrow} 0,\quad
\left . \begin{array}{c}A\emptyset \\ \emptyset A \end{array}
\right \} @>w (1-q)/2>> AA,
\end{CD}
\label{Eq:cross}
\end{equation}
where $0 \le q \le 1$.
If $w \neq 0$, the system shows the DP scaling behavior~\cite{PP2006}.

What can we expect about the scaling behavior near the PCPD
critical point for finite $w$ if PCPD does belong to the DP class?
From the rigorous study about the ``crossover''\footnote{We use
the quotation mark to mention the ``crossover'' between models in
the same class because the ``crossover'' is not truly a crossover
in convention.}  model of  the BAW with one
offspring~\cite{PP2007} based on the stochastic equivalence shown
in~\cite{PJP2005}, the ``crossover'' between the same universality
classes is expected to have two characteristics. First, the phase
boundary is expected to meet the PCPD point linearly. Second, the
critical amplitudes which is defined as $\rho(t)
t^{\delta_\text{DP}}$ for cases with finite but small $w$ should
collapse with the critical amplitude of the PCPD at $w=0$. Our
recent numerical results show a nontrivial crossover exponent from
the PCPD to the DP and the phase boundary approaches the PCPD
point in a singular (nonlinear) way~\cite{PP2006}. This provides
another strong evidence that the PCPD is different from the DP.

\begin{figure}[t]
\resizebox{\columnwidth}{!}{
\includegraphics{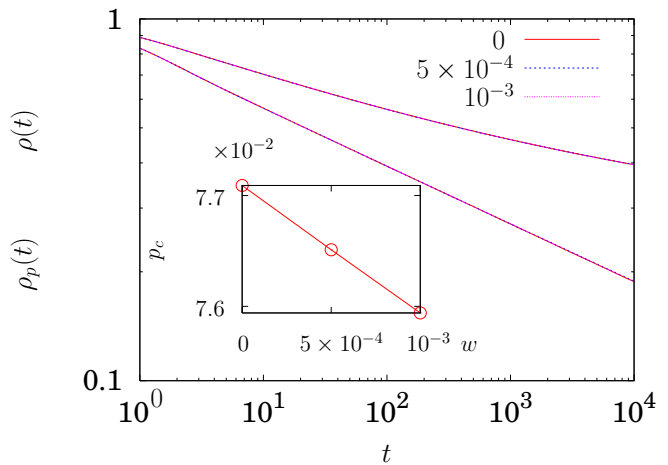}
}
\caption{\label{Fig:pcpima} (Color online) Log-log plots of the particle density (upper
curves) and
the pair density (lower curves) at criticality vs $t$ for the PCP
``crossover'' model
with the dynamics $3A \rightarrow 0$ which occurs
with probability $w$. The critical points are
$0.077~0905(5)$, $0.07651(1)$, and $0.07594(1)$ for $w=0$, $w=5\times 10^{-4}$,
and $w=10^{-3}$, respectively.
The difference among cases with $w=0$ and finite $w$ is hardly seen.
Inset: The phase boundary near $w=0$. The line meets the vertical
axis linearly.
}
\end{figure}

In fact, one should be cautioned in interpreting the singular
behavior of the phase boundary. The linearity of the phase
boundary may become complicated by the nontrivial singularity
arising between the PCP and the DP crossover~\cite{PP2007}, which
corresponds to the dynamics modeled by Eqs.~\eqref{Eq:PCPDrule}
and \eqref{Eq:cross} with $D=0$. However, this singularity has
nothing to do with the universality class. In a sense, the model
with $w=0$, i. e., the PCP, is pathological in that the
configuration volume occupied by the absorbing states is
macroscopic, which is not the case for finite $w$\footnote{When
$w$ is finite, the only absorbing state is the particle vacuum.}.
Hence there is an inherent singularity, actually discontinuity of
the particle (auxiliary field) density, close to $w=0$, which is
reflected by the nontrivial crossover exponent although the PCP
belongs to the DP class~\cite{PP2007}. If we introduce $3A
\rightarrow 0$ rather than Eq.~\eqref{Eq:cross}, we reproduced two
characteristics of the ``crossover'' between models in the same
universality class; see Fig.~\ref{Fig:pcpima}. The results
summarized in Fig.~\ref{Fig:pcpima} are rather easily conceivable
because the operator corresponding to $3A \rightarrow 0$ is
irrelevant in the RG sense and moreover there is no singularity of
the auxiliary field density near $w=0$. In this context, the
single particle branching/annihilation introduced to the PCP is
relevant because this operator changes the structure of the
absorbing configurations, which is manifest by the singularity at
$w=0$~\cite{PP2007}, though it does not change the universality class.

Since the volume of the absorbing states in the configuration
space for the PCPD is zero in the thermodynamic limit, there is no
singularity arising from the auxiliary field near the crossover
from the PCPD to the DP. Hence the reason of the singularity in
the PCPD-DP crossover should be understood in another context. The
easiest answer may be that the PCPD does not belong to the DP. Of
course, there could be a hidden unknown reason for the PCPD to
lead to a singular crossover behavior and  its possibility cannot
be fully excluded. Hence we studied a similar model to the PCPD
whose non diffusing counter part has infinitely many absorbing
states and which is known to belong to the DP class. It is found
that the crossover model based on the \texttt{tp12} have two
properties of the ``crossover'' among models belonging to the same
universality class~\cite{PP2007}.
We are unable to conceive of a mechanism that would explain these
observations, while maintaining the PCPD and \texttt{tp12} in the
same universality class; a more natural interpretation is that the
two models belong to different classes, so that the PCPD does not
fall in the DP class. All together, our study of the crossover
scaling strongly suggests that the PCPD does not belong to the DP
class.

\section{\label{Sec:con}Conclusion}
Up to now, we discussed about the hotly debated issue of the pair
contact process with diffusion (PCPD). Among many scenarios
proposed at the first stage of the controversy~\cite{HH2004}, only
two seem to have survived; does the PCPD belong to the directed
percolation class or form a different universality class from any
other known one? In this paper, we gave evidences in favor of the
second scenario.

At first, the fact that the biased diffusion drastically changes
the critical behavior of the PCPD suggests that the one
dimensional PCPD should be described by the two independent
relevant ``elementary excitations'' which is not the case of the
Reggeon field theory (DP). Second, the nontrivial crossover
scaling arising between the PCPD and the DP strongly suggests that
the PCPD should not belong to the DP class.

Although the main interest of this paper is the PCPD, the methods
employed to clarify the universality issue of the PCPD (the role
of the biased diffusion and the crossover) are generally
applicable to many other systems, some of which are currently
under investigation.


\end{document}